\documentclass[aps,pre,10pt,notitlepage]{revtex4-2}

\usepackage{amsmath,amssymb,amsfonts,mathrsfs,graphicx,xcolor}

\begin{document}

\title{Self-driven criticality in a stochastic epidemic model}

\author{Gil Ariel} \affiliation{Department of Mathematics, Bar-Ilan University, Ramat Gan 52000, Israel}

\author{Yoram Louzoun} \affiliation{Department of Mathematics, Bar-Ilan University, Ramat Gan 52000, Israel}

\begin{abstract}

We present a generic epidemic model with stochastic parameters, in which the dynamics self-organize to a critical state with suppressed exponential growth. More precisely, the dynamics evolve into a quasi-steady-state, where the effective reproduction rate fluctuates close to the critical value one for a long period, as indeed observed for different epidemics. The main assumptions underlying the model are that the rate at which each individual becomes infected changes stochastically in time with a heavy-tailed steady state. The critical regime is characterized by an extremely long duration of the epidemic. Its stability is analyzed both numerically and analytically in different models.

\end{abstract}

\maketitle

\section{Introduction}

Compartmental epidemics models, and their many variations or derivatives have been proven useful in understanding, analyzing and predicting real epidemic outbreaks \cite{Herhcote2000,Murray2007}. 
However, fitting such models to the observed infection or death time series has proven challenging \cite{Akira2020}. 
One of the main discrepancies is that compartmental models are essentially exponential, at least locally in time \cite{Rhodes1997,Herhcote2000}, while observed data are often not \cite{BenNaim2004,Murray2007,Radicchi2020}. 
Exponential dynamics emerge since at any given time in the evolution of the epidemics, the equation for the dynamics can be linearized around its current state, suggesting an exponential growth or decay of variables (except at particular time points such as the local maximum of infected individuals). 
The predicted exponential growth/decay  motivates the commonly used notion of the effective reproduction rate, $R_t$, or equivalently, doubling time \cite{Herhcote2000,Murray2007}. 
The effective reproduction rate describes approximately the instantaneous exponential rate of change in the number of infected, hospitalized, deceased or other types of individuals. 
Fitting $R_t$ to real data is not straight-forward, and several methods have been proposed and applied \cite{Hotz2020,Junling2020,Manrique-Abril2020}.

Often, the epidemic dynamics seem to be at, or close to the critical state $R_t=1$ \cite{Grassberger1983,Rhodes1997,Stollenwerk2003,Stollenwerk2005,Ion2017,Brett2020,Contoyiannis2020,Gans2020}. 
Consequently, the number of new cases per day is constant or linear. 
Indeed, several authors studied the dynamics of epidemics, both in compartmental and network models, assuming that the epidemic is poised at the critical threshold between exponential growth and exponential decay \cite{Grassberger1983,BenNaim2004,Horstmeyer2018,Radicchi2020}.
This dynamical pattern can be explained by assuming that contact rates adapt to 
the spread of the epidemic to fine-tune $R_t$ \cite{Brett2020,Gans2020}. 
However, such negative feedback typically takes effect on long time scales, possibly up to years \cite{Brett2020}.
Alternatively, Stollenwerk and Jansen \cite{Stollenwerk2003,Stollenwerk2005} suggested a sand-pile-type model that exhibits self-organized criticality. 
The model assumes that the epidemic spreads on a square 2D lattice. 
Criticality is due to a vanishingly small rate of mutation to a deadly strain. 
Recent versions assume that the epidemic spreads to neighbors only once the viral load is above a threshold \cite{Contoyiannis2020} or that lattice sites are partially isolated communities (cliques) \cite{Ion2017}. 
Although these models are applicable for specific examples  \cite{Stollenwerk2003,Stollenwerk2005}, they are not as generic as standard compartmental models. Moreover, there is no evidence supporting their main assumptions
in general.

Here, we suggest a generic mechanism driving epidemics towards criticality.
The main new assumption is that the infectivity of each individual (i.e. the probability of each individual to get infected upon the meeting with an infected individual) is a time-varying correlated random variable. 
Three dynamical regimes or phases are identified, depending on parameters: (i) Exponential growth, (ii) Exponential decay, and (iii) a quasi-steady critical state - A novel regime in which the dynamics naturally evolves to a steady-state that is close to critical, i.e., $R_t$ fluctuates close to 1. 
In this regime, the exponential dynamics are suppressed, resulting in constant or linear infected individuals over long time scales. The critical state does not require fine-tuning of parameters to have particular prescribed values.
The only condition is for a heavy tail in the infectivity probability distribution
and a finite correlation time.

\section{Model details}

For simplicity, we concentrate on the SIR modeling point of view, assuming that the evolution of the disease in each individual follows the three states (S)usceptible -> (I)nfected -> (R)ecovered \cite{Herhcote2000,Murray2007}. 
In addition, we assume a well-mixed population \cite{Herhcote2000}. These assumptions underlie considerable simplifications, and realistic epidemic modeling requires more detailed models \cite{Herhcote2000,Epstein2009,Britton2020}. 
However, our main goal is to show that the new self-organized critical regime is generic. 
Hence, we restrict our focus to the simplest, most universal approach. The mechanism proposed here applies similarly to more complex models. Similarly, for the sake of simplicity, we assume that the variability is only on the infected side, and not on the infecting side.

The SIR model dynamics are determined by three parameters: The population size $N$, the infectivity rate $\lambda$, and the recovery rate, taken with no loss of generality to be 1. 
Our only departure from the standard SIR model is in relaxing the assumption that the infectivity rate of an individual is constant. Instead, we assume it follows a stochastic process, i.e., it is changing randomly in time. 
In other words, the infectivity rate of individual $k$ is replaced by a stochastic process $\lambda_k (t)$.  
Thus, each infected individual can infect each susceptible individual $k$ with a rate proportional to the susceptible infectivity, $\lambda_k(t)$.

Different biological and social mechanisms can be proposed to explain a random evolution of the individual susceptible infectivity rate. 
For example, individual differences in health, behavior, social distancing measures, and more. Indeed, previous models considered the effects of quenched heterogeneity (i.e., assume that individuals have different properties, but they are fixed in time) \cite{Miller2007,Louzoun2016,Britton2020,Louzoun2020}. 
Several authors considered deterministic time dependence of $\lambda_k (t)$, for example due to seasonality or decline in vaccination rates \cite{Kloeden2011,Kloeden2015,Brett2020}.  Stochasticity has also been considered, either through a randomly evolving network \cite{Ochab2011}, by additive and multiplicative noise \cite{Tornatore2005,Jiang2011,Kloeden2015,Caraballo2017} or by stochastic (typically normal) fluctuation of parameters \cite{Dureau2013,Faranda2020}. 
The main conclusion of these approaches is that assuming population growth (possibly with a random but strictly positive growth rate), stochastic SIR models admit a steady-state solution. 
This solution does not exist if the population size $N$ is fixed. 
Intuitively, the idea underlying these models is clear: at the steady-state, the average population growth compensates for the average rate of new infections. This steady-state is reached on the time scale of the population growth, typically considerably longer than the epidemic time scale. 

Here, we assume that infectivity rates $\lambda_k (t)$ are independent, identical stationary stochastic processes with two important properties: (i) univariate marginal distributions that have a power-law tail, (ii) a finite correlation time. 
Note that, while individual infectivity rates vary, the statistics of all individuals are identical and constant in time.  In particular, the process is autonomous, i.e., there is no explicit time dependence. 
To be specific, the model is defined as follows. 

Below, we denote the fraction of Susceptible, Infected and Recovered populations by $s(t)$, $i(t)$ and $r(t)$, respectively, with $s(t)+i(t)+r(t)=1$. 
A susceptible individual $k$ becomes infected in the period $[t,t+dt)$ with probability $p_k ({\rm S} \to {\rm I})= \lambda_k (t) i(t) dt+o(dt)$. 
An infected individual $k$ recovers in the period $[t,t+dt)$ with probability $p_k ({\rm I} \to {\rm R})=dt+o(dt)$. 
There are different ways to implement a stochastic process satisfying the two conditions described above.
We choose several realizations, and present here one that is convenient to simulate and admits some analytical analysis.
The main idea is to define the characteristic infectivity times $\mu_k(t)=1/\lambda_k(t)$ instead of the rates.
We assume that $\mu_k(t)$ are independent stationary processes with gamma univariate marginal distribution,
\begin{equation}
   \mu_k(t)\sim {\rm Gamma} (\alpha,\beta)
\label{eq:ss1}
\end{equation}
and an exponential auto-correlation function 
\begin{equation}
   C(t,s)=\frac{\beta^2}{\alpha} \left[ \left< \mu_k(t) \mu_k(s) \right> - \left< \mu_k(t) \right> \left< \mu_k(s) \right> \right] = e^{-|t-s|/\tau} .
\label{eq:ss2}
\end{equation}
Then, the instantaneous distribution of infectivity rates $\lambda_k(t)=1/\mu_k(t)$ has an inverse gamma distribution, characterized by a tail that decays with a power $\alpha+1$.

Writing the SIR model as a continuous-time Markov process, we simulate the dynamics using the Gillespie algorithm \cite{Allen2017}. Synchronous simulations give similar results.
We consider shape parameter in the range $1<\alpha<2$, which implies that infectivity rates have a finite mean, but infinite variance. The rate parameter $\beta$ is chosen to have a given average transition time $\left< \mu_k(t) \right>=\alpha/\beta$.
Initial conditions are 10 infected individuals and no recovered.   
See the appendix for implementation details.

\section{Simulation results}

\begin{figure}[]
\includegraphics[scale=0.6, trim=30 75 100 10, clip]{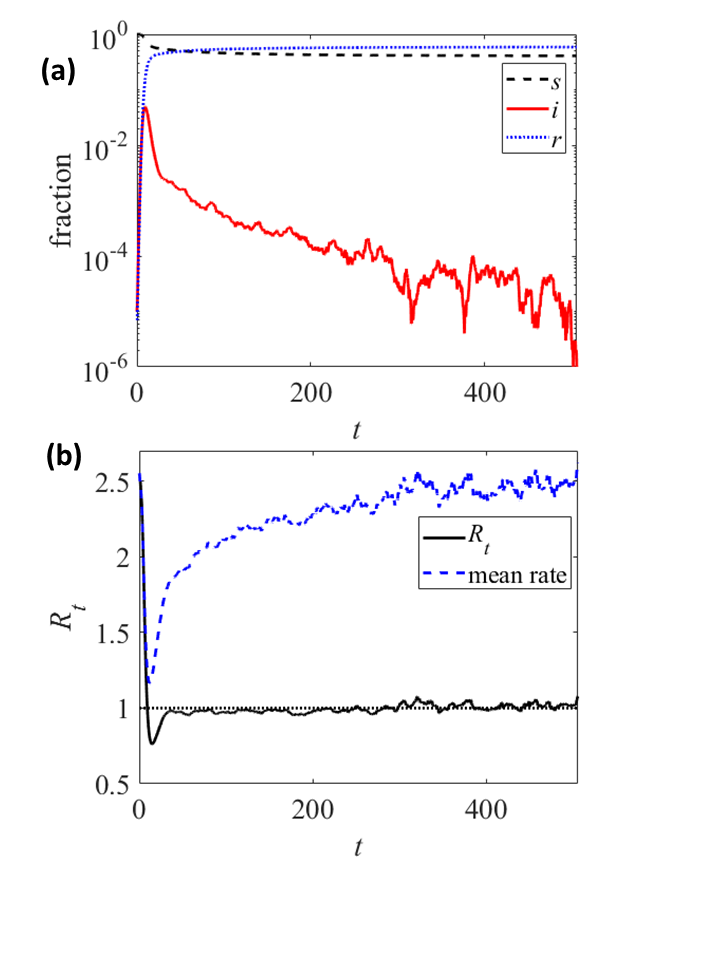}
\caption{\label{fig1} 
Example simulation results with a population of 1M individuals, average infectivity rate 2.7 and shape parameter $\alpha=1.3$. Time units are the inverse rate of recovery. (a) The time evolution of the fraction of susceptible, $s(t)$, infected, $i(t)$ and recovered $r(t)$ individuals on a semilog scale. (b) After a short relaxation period, the effective reproduction rate $R_t$ fluctuates close to the critical value 1. As $R_t$ remains constant while $s$ decreases, the average infectivity rate has to increase.}
\end{figure}

Fig.~\ref{fig1}a shows the time evolution of $s$, $i$ and $r$ for a population of $10^6$ individuals. 
Parameters are $\alpha=1.3$, $\left< \mu_k(t) \right>=1.6$ (i.e., $\beta\simeq 0.81$ and $\left< \lambda_k(t) \right> \simeq 2.7$) and $\tau=4$.
Following a short exponential transient that follows the SIR dynamics with either equal or quenched infectivities,  most individuals that started with a high infectivity rate are infected, and the rate of new infections decreases (fig.~\ref{fig1}b). 
As a result, the fraction of infected individuals decreases, but on a considerably longer time scale compared to previous models with non-critical dynamics. 
Fig.~\ref{fig1}b depicts our main result, showing that the effective reproduction rate $R_t$ stabilizes to a value that fluctuates around the critical value $R_t=1$. 

An intuitive explanation is a  negative feedback mechanism: 
Initially, the average infectivity rate is larger than 1. 
As $s$ is close to 1, the reproduction rate $R_t>1$ and the number of infected individuals increases exponentially.  
However, susceptible individuals with large initial  $\lambda_k (0)$ are more likely to be infected during this exponential growth phase.  
As a result, given the heavy tail of the distribution of $\lambda_k$, the average infectivity rate in the remaining susceptible population decreases, decreasing $R_t$. 
On the other hand, if the rate of new infections $i'(t)$ is sufficiently large (compared to $1/\tau$) so that $R_t<1$, then the number of infected individuals decreases exponentially. 
Since there are not many new infections, and because the distribution of infectivity rates returns to the steady-state at a characteristic time $\tau$, the average infectivity rate increases back towards its unperturbed steady-state average, which is greater than one. Note that an essential requirement for such a mechanism is a heavy-tailed distribution of the infectivities, otherwise, the reduction of the right-hand tail would not have a strong affect on the average infectivity. In addition, if the correlation time $\tau$ is too small or large, then the infectivity rate will not return towards its steady-state value. 

Overall, if the increase and decrease in the average infectivity rate occur on comparable time scales, then the dynamics self-organize to the critical value. 
In the SIR model, the fraction of susceptible individuals $s(t)$ decreases monotonically with time.
Denoting averages over susceptible individuals by $\langle \cdot \rangle_{\rm S}$, the instantaneous reproduction
rate is given by $R_t=\langle \lambda_k (t) \rangle_{\rm S} s(t)$.
Thus, criticality is maintained through an increase in the average infectivity of each susceptible, through a stronger bias for infections in the tail of the distribution (dashed blue curve in Fig.~\ref{fig1}b).

\begin{figure}[]
\includegraphics[scale=0.8, trim=0 240 0 0, clip]{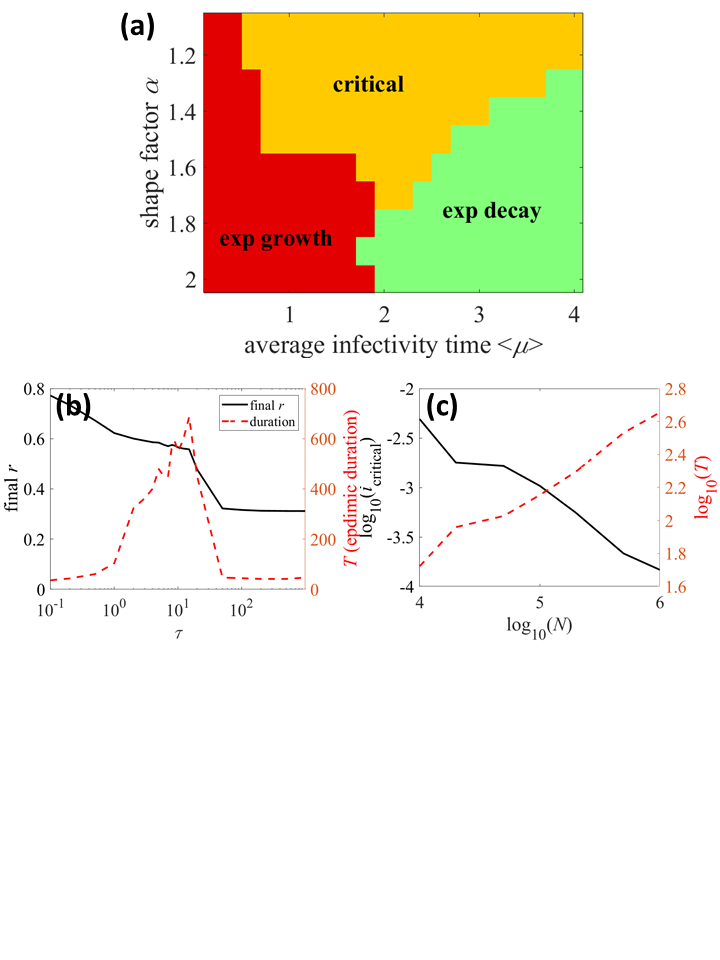}
\caption{\label{fig2} 
phase diagram. (a) The regime of the epidemic (exponential growth, exponential decay, or self-organization to the critical state) for different average infectivity times and shape factor $\alpha$. The critical regime occupies a large fraction of the phase space and does not require fine-tuning of parameters.
(b) Dependence on $\tau$. The figure shows the final values of $r$, indicating the overall fraction of the population which was infected (solid black line),
and the duration of the epidemic (dashed red line) as a function of $\tau$.
As the critical state is characterized by long epidemic duration, we see that
the stability of the $R_t=1$ state increases for $\tau \ge 1$ and abruptly drops around $\tau=20-50$, suggesting a phase transition.
(c) Dependence on $N$. The figure shows $i_{\rm critical}$, the fraction of the population infected during the critical phase (solid black line),
and $T$, the epidemic duration on a log-log scale (dashed red line). 
Initial conditions are $I(0)=10^{-4} N$ and results are averaged over 10 realizations. 
Both quantities seem to depend algebraically on $N$, with slopes -0.8 and 0.5, respectively. }
\end{figure}

\begin{figure}[]
\includegraphics[scale=0.8, trim=195 50 190 40, clip]{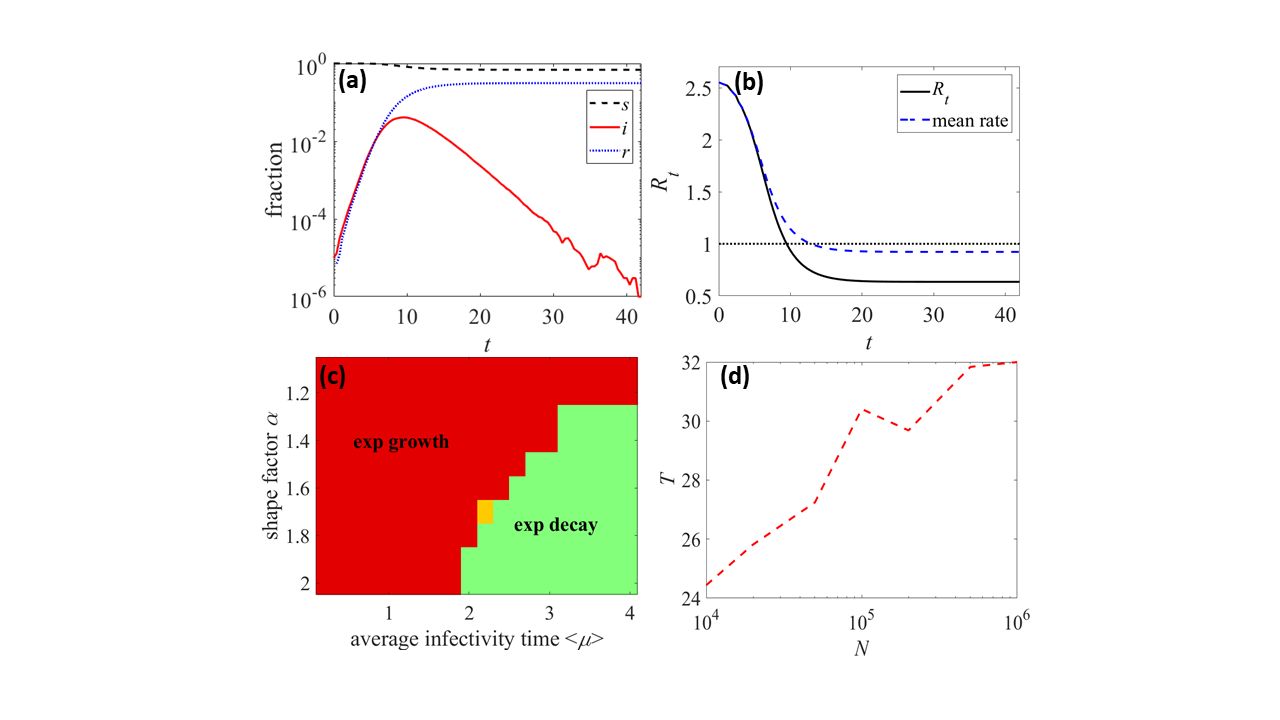}
\caption{\label{fig3} 
Results for annealed dynamics, $\tau \to \infty$. All other parameters are the same as in fig.~\ref{fig1}. (a) $s(t)$, $i(t)$ and $r(t)$ on a semilog scale. The duration of the epidemic is an order of magnitude shorter compared to the critical regime. Time units are the inverse rate of recovery. (b) Both the reproduction number $r(t)$ and the average infectivity rate decrease monotonically in time. (c) The phase diagram shows that criticality is only obtained at the exact boundary between the exponential growth and exponential decay regimes.
(d)  Dependence on $N$. The duration of the epidemic $T$ appears to increase sub-logarithmically with $N$.
}
\end{figure}

The duration of the critical state can be orders of magnitude longer than in the quenched analog (Fig.~\ref{fig3}). 
However, it does not correspond to a true steady-state due to the finite population.
Accordingly, we term this regime a quasi-steady critical state.
To understand the different regimes that emerge in this model, we estimate the model phase diagram. 
Fig.~\ref{fig2}a shows the three dynamical regimes (exponential growth, exponential decay or quasi-steady criticality) as a function of the average infectivity time and the shape parameter. 
We refer to the exponential decay case as an instance in which the overall (final) fraction of the recovered population (when the infected population reaches 0) is smaller than
a low threshold (here we used 1\%).  
Otherwise, the epidemic spreads (with exponential or critical growth).  
The regime is labeled critical if $R_t$ is in in $[1-\epsilon,1+\epsilon]$, for at least half of the duration of the epidemic.
Here we used $\epsilon=0.05$.
Otherwise, the regime is labeled as exponential growth. 
Note that these definitions contain multiple arbitrary parameters, which may affect the precise borders of the different regimes. However, the main properties of the phase diagram, 
in particular the occurrence of the critical regime in a wide range of parameters, 
are not affected by the details of the formal definitions.
For sufficiently large vales of $\alpha$, the boundary between the exponential growth
and decay regimes occurs when $\left< \mu \right>=2$, in which $R_t$ is approximately 1.

Fig.~\ref{fig2}a shows that the critical regime inhabits a large section of the phase diagram. 
Not surprisingly, it is highly dependent on $\alpha$.
When $\alpha$ approaches 1, the dynamics are critical for a very wide range of beta values.
For high $\alpha$ values, the model converges to the classical SIR model distinction of exponential increase for high infectivities (lower time for infection), and exponential decrease for low infectivities. 
Fig.~\ref{fig2}b shows the dependence of the dynamics on the characteristic time-scale of the stationary gamma process, $\tau$.
All other parameters are the same as in fig.~\ref{fig1}.
The final value of $r$, indicating the overall fraction of the population which was infected,
decreases monotonically with $\tau$. 
However, the duration of the epidemic depends non-monotonously on $\tau$.
As the critical state is characterized by long epidemic duration, the figure shows that
the stability of the $R_t=1$ state increases for $\tau \ge 1$ and abruptly drops around $\tau=20-50$. 
The sharpness of the drop in the duration suggests a phase transition.
Finally, we study the dependence on the population size $N$.
To this end, systems with different population sizes were simulated with 
equal initial condition $i(0)=10^{-4}$. 
The epidemic duration $T$ was defined as the last time where $i(t)>=10^{-5}$.
The fraction of infected individuals during the critical state was defined as $i_{\rm critical} = i(T/2)$.
Both $i_{\rm critical}$ and the epidemic duration $T$ show an algebraic dependence on $N$, with slopes of approximately -0.8 and 0.5, respectively (Fig.~\ref{fig2}c).
The duration of the quasi-steady state increases with $N$, suggesting it may become a fixed-point in the infinite system limit ($N \to \infty$).

These results are in stark contrast to results is quenched infectivity rates (Fig.~\ref{fig3}).
In the quenched system, the dynamics transitions from early exponential growth to exponential decay, with $R_t$ monotonically decreasing to an asymptotically value that smaller than 1. In particular, the critical state $R_t=1$ is instantaneous.
Accordingly, the phase diagram shows that criticality is only obtain for specific fine-tuned values along the boundary between the exponential growth and decay phases.
Finally, the epidemic duration decays sub-logarithmically with $N$ (Fig.~\ref{fig3}d). 
This is due to the fact that $T$ was calculated until a fixed small fraction $i(t)$ is reached.
With $R_t<1$ (exponential decay), the average time from $i(t)>=10^{-5}$ to $i(t)=0$ is trivially of order $\ln N$. 
Overall, the quenched model does not show a critical behavior.

\begin{figure}[]
\includegraphics[scale=0.7, trim=0 80 10 190, clip]{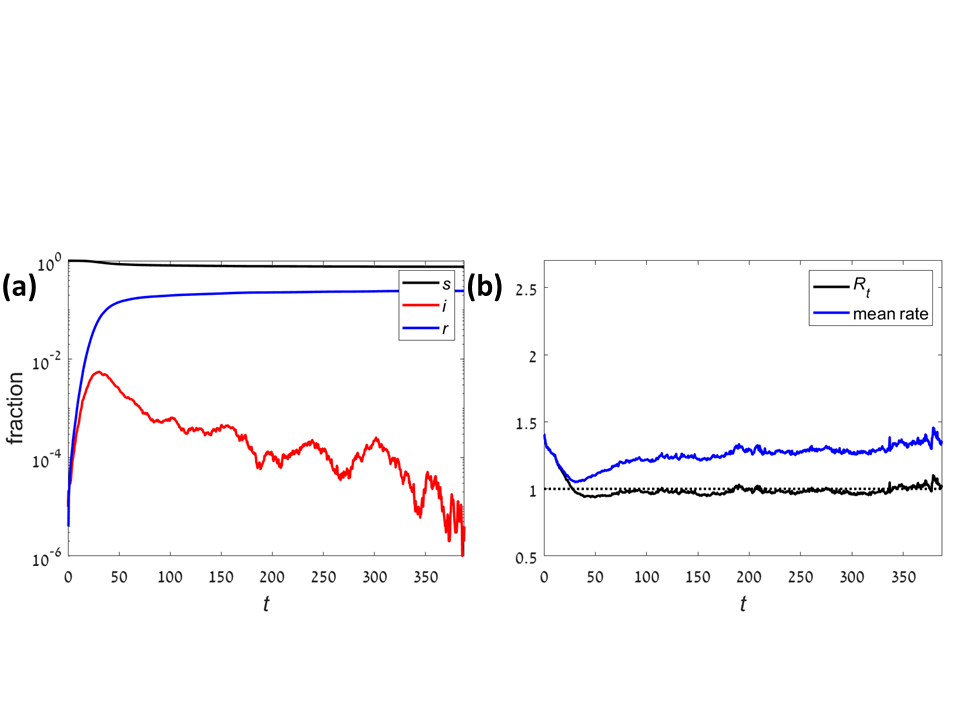}
\caption{\label{fig4} 
Simulation results with the alternative jump rates processes. $\alpha=1.3$, $\left< \lambda \right> = 1.4$, $\tau=4$ and $N=10^6$.
Time units are the inverse rate of recovery. (a) The time evolution of the fraction of susceptible, $s(t)$, infected, $i(t)$ and recovered $r(t)$ individuals on a semilog scale. (b) After a short relaxation period, the effective reproduction rate $R_t$ fluctuates close to the critical value 1. As $R_t$ remains constant while $s$ decreases, the average infectivity rate has to increase. 
}
\end{figure}

The appendix describes an alternative implementation for simulations, in which
the infectivity rates are jump processes with univariate marginal distributional that are ``pure'' power-laws.
The results, depicted in Fig.~\ref{fig4}) are qualitatively similar to Fig.~\ref{fig2}, showing a long quasi-steady state.

\section{An analytical approximation of the critical regime}

\begin{figure}[]
\includegraphics[scale=0.7, trim=0 0 10 0, clip]{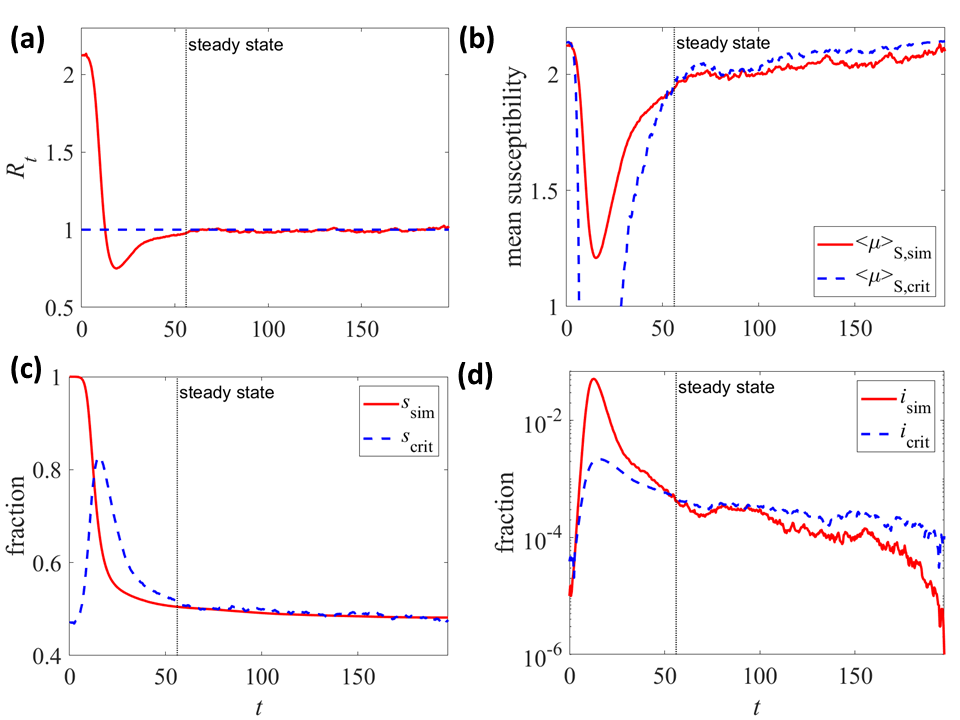}
\caption{\label{fig5} 
Comparison between simulation results with $\alpha=3/2$, $\left< \mu \right> = 1.4$, $\tau=4$ and $N=10^6$ and the analytical approximation at the steady state.
(a) After an initial transient period, the reproduction rate $R_t$ fluctuates close to 1.
The dashed vertical line marks the beginning of the quasi-steady state in simulations.
The agreement between the analytical and simulated results slowly deteriorates because in simulations $S$ is slowly decreasing and
the steady-state assumption does not hold.
}
\end{figure}

To further gain insight into the stability of the critical regime,   
we take advantage of a particular realization of stationary gamma processes for the case in which the shape parameter $\alpha$ is equal to half an integer.
From \cite{Wolpert2011,Wolpert2016}, the solution to the following stochastic differential equation,
\begin{equation}
   d\mu_t = -\frac{2}{\tau} \left( \mu_t - \frac{\alpha}{\beta} \right) dt + \frac{2}{\sqrt{\beta \tau}} \sqrt{\mu_t} dW_t ,
\label{eq:SDE}
\end{equation}
satisfies (\ref{eq:ss1}) and (\ref{eq:ss2}), where $W_t$ is the standard Brownian motion. 
In order to derive self-consistent equations for the existence of a steady-state,
we assume that the total rate of individuals leaving state S (i.e., they become infected) is constant (that depends on $i$).
In other words, $\mu_t$ is a diffusion process that is killed at a rate that is proportional to $\mu_t^{-1}$.
To uphold the steady-state, individuals are reintroduced to the susceptible population at a constant rate.
One can write the Fokker-Plank equation associated with the effective steady-state distribution $p(\mu)$ (see the appendix).
Expanding in an asymptotic series,
\begin{equation}
   p(\mu) = p_0 (\mu) + i_{\rm critical} p_1(\mu) + O(i_{\rm critical}^2) ,
\end{equation}
where $p_0(\mu)$ is the density of the Gamma distribution ${\rm Gamma} (\alpha,\beta)$ and $i_{\rm crit}$ is the fraction of infected individuals during the steady-state.
In the appendix, we show that $q(\mu)=2p_1(\mu)/\tau$ is dimensionless, and satisfies an inhomogeneous confluent hypergeometric equation
whose solutions can be written in terms of a generalized gamma function. 
With $\alpha=3/2$, we find that a steady state can be obtained for a large range of parameters with a critical fraction of infected individuals that is inversely proportional to the population size, $i_{\rm critical}$ is at most $1/\ln N$.
Comparison with the numerical results confirm the stability of the critical state and the parameters range in which it is obtained (Fig.~\ref{fig5}).
In contrast, for $\alpha=2$ and $\alpha=5/2$, a steady-state cannot be obtained for long durations that are at least of order $\ln N$. 
Therefore, the phase diagram (Fig.~\ref{fig2}b) can be understood by the following logic.
If the average infectivity rate $\left< \lambda \right>$ is too small (i.e. $\left< \mu \right>$ is large), then the initial $R_t<1$ and we observe an exponential decay.
On the other hand, if $\left< \lambda \right>$ is very large, $i(t)$ becomes
too large ($O(1)$ in $N$) and the quasi-steady state is not possible (or short).
In between, critical dynamics is observed.

\section{Discussion}

Some fundamental questions are still open. 
The generality of our results remains to be proven beyond the three examples studied here.
In addition, linking the self-organizing principle to similar known physical processes, in particular the well-studied self-organized criticality \cite{Stollenwerk2003,Stollenwerk2005,Contoyiannis2020,Ion2017} and the stable steady-state inferred in stochastic models (with population growth) \cite{Tornatore2005,Jiang2011,Kloeden2011,Ochab2011,Kloeden2015,Brett2020,Caraballo2017} or to the steady state with an effective $R_t=1$ observed in SIRS models, is of interest.

Relating the model proposed here to realistic epidemic data requires introducing more complicated models to take into account several details that have been proven important to realistic modeling \cite{Wearing2005}, such as interaction networks,
heterogeneity (e.g. by age), incorporating the variability in the infecting side, and coupling between infectivity rates and the epidemic dynamics. 
Most importantly, the current work does not establish that the main assumption underlying the model are indeed realized for some detailed, microscopic model. For example, it has been suggested that the rate of some person-to-person contact types were are heavy-tailed \cite{Eubank2004,Manzo2020} (while others are not \cite{Liljeros2001}). 
In addition, some contact durations, which may also influence infectivity rates, were also found to be heavy tailed \cite{Montanari2017}. 
SIR-type models are also widely used to describe the spreading of computer viruses \cite{Louzoun2020}, memes in social networks \cite{Yang2018} and adoption of new products \cite{Fibich2016}. 
It is well known that the number of edges in such networks is heavy-tailed \cite{Pastor2001,Louzoun2020}. 
These issues are beyond the scope of the current manuscript.

To summarize, we showed that a small change in the SIR model in which the infectivity rates (or rather times) of each individual are changing stochastically in time can qualitatively change the dynamics. 
For simplicity, we assumed a minimal model with a well-mixed population, statistically identical individuals, and a stationary, independent random process for the infection time of each individual. Our results demonstrate that the effect of including temporal fluctuations in the infectivity times is drastic. In particular, the dynamics admit a new regime that is not present in previous compartmental models, in which the dynamics evolve spontaneously into a quasi-steady critical state in which the reproduction number fluctuates close to 1. 
During this stage, a small, constant fraction of the population, $i_{\rm critical}$, remains infected. This fraction decreases with $N$. 
At the same time, the duration of this state increases with the population size approximately as $\sqrt{N}$, suggesting a phase transition
to a critical steady-state in the infinite system limit.

Without a doubt, fitting and understanding realistic epidemic data require further generalizations to more complicated models. 
Yet, this work takes the first steps in understanding the implications of the self-organized quasi-stable critical regime introduced here, which can have analog consequences in other systems, such as population dynamics and kinetics of chemical reactions.

{\bf Acknowledgments} We thank Rainer Klages and Jeremy Schiff for comments and discussions. 
We thank Robert Wolpert for sharing his notes on stationary gamma processes.


%
%


\appendix
 \section{Appendix: Simulation implementation}

The section provided details for implementing the SIR simulations described in the main text.
Recall that for each $k$, $\mu_k(t)$ is a stationary process with univariate marginal distribution $\mu_k(t) \sim {\rm Gamma} (\alpha,\beta)$
and exponential auto-correlation function $\exp(-|t-s|/\tau)$ for $\mu_k(t)$ and $\mu_k(s)$.
The initial distribution of infection times is taken to be the steady-state distribution ${\rm Gamma} ( \alpha,\beta)$. Thus, the process for $\mu_k (t)$ is stationary. 

We approximate $\mu_k(t)$ by a piece-wise constant jump process with step $\Delta t$ using the discrete AR(1) process
presented in [37-39]. 
To be precise, let $\rho=\exp(-\Delta t/\tau)$ and take
\begin{equation}
   \mu_k ((i+1)\Delta t) = \rho \mu_k(i\Delta t) + \zeta_i ,   
\label{eq:AR1}
\end{equation}
where,

\begin{eqnarray}
   G_i &\sim & {\rm Gamma} (\alpha,1) \\
   N_i &\sim & {\rm Pois} ((1-\rho)G_i/\rho) \\
   \zeta_i &\sim & {\rm Gamma} (N_i,\beta/\rho) .
\end{eqnarray}
Here, ${\rm Pois} (L)$ denotes a Poisson distribution with mean $L$.

The SIR model is simulated in steps of $\Delta t$.
During the $i$'th step, we assume that S$\to$I transitions have an exponential distribution with constant rates $\lambda_k=1/\mu_k(i\Delta t)$.
I$\to$R transition rates equal 1.
Within each time period $\Delta t$, the SIR dynamics is modeled using the Gillespie algorithm as follows.
For simplicity, we consider the first period, $i=0$ and $t=0$.
Denote by $A_{\rm S} (t)$ and $A_{\rm I} (t)$ the sets of susceptible and infected individuals at time $t$, respectively. 
\begin{enumerate}
\item
   Let 
   \begin{equation}
      \Lambda = \frac{1}{N} \left| A_{\rm I} (t) \right| \sum_{k \in A_{\rm S} (t)} \lambda_k ,
    \end{equation}
    where $\left| A \right|$ denotes  the number of elements in a set $A$.
\item
   Let $dt_{\rm S2I}$ denote the time to the first infection event. As the rate in which a susceptible individual $k$ becomes infected is $i \lambda_k/N$, 
   the first event occurs at
   $dt_{\rm S2I} \sim {\rm Exp} (1/\Lambda )$, an exponentially distributed random variable with mean $1/\Lambda$.
\item
   Let $dt_{\rm I2R}$ denote the time to the first recovery event. As the rate in which an infected individual recovers is 1, 
   the first event occurs at
   $dt_{\rm I2R} \sim {\rm Exp} (1/\left| A_{\rm I} \right|)$.
\item   
   Let $dt = \min \{ dt_{\rm S2I},dt_{\rm I2R} \}$.
\item
   If $t+dt>\Delta t$, then no transitions occur until the end of the period $[0,\Delta t]$ and the process jumps to $t=\Delta t$. The current time-step ends. 
\item
   Otherwise, a transition occurs. 
   If $dt_{\rm I2R} < dt_{\rm S2I}$, then a recovery event occurs. 
   Let $I$ denote an element from $A_{\rm I} (t)$ chosen at random with uniform distribution. Then, individual $I$ recovers:
   $A_{\rm I} (t+dt) \leftarrow A_{\rm I} (t)-\{I \}$ and $t \leftarrow t+dt$. Return to 1.
\item
   Otherwise, an infection event occurs. Let $u \sim U(0,1)$, a uniformly distributed random variable in $(0,1)$, and 
   \begin{equation}
      I = \arg \max_I \left\{ \frac{ \sum_{k \in A_{\rm S} (t),k\le I} \lambda_k }{ \sum_{k \in A_{\rm S} (t)} \lambda_k } < u \right\} .
   \end{equation}
   Then, individual $I$ becomes infected:
   $A_{\rm S} (t+dt) \leftarrow A_{\rm S} (t)-\{I \}$, $A_{\rm I} (t+dt) \leftarrow A_{\rm I} (t) \cup \{I \}$ and $t \leftarrow t+dt$. Return to 1.
\end{enumerate}
Note that the algorithm makes use of the no-memory property of exponential random variables. Thus, cutting off the distribution for $dt$ if $t+dt>\Delta t$ and redrawing it does no change the distribution of $dt$ (unless $\Lambda$ or $ \left| A_{\rm I} (t) \right|$ change).

 \section{Appendix: Infectivity rates as a jump process}

The section describes an alternative mechanism that does not use stationary gamma processes.
The main idea is to define a stochastic process for the infectivity rates $\lambda_k (t)$ such that
\begin{enumerate}
\item The univariate distribution of $\lambda_k(t)$ are a power-law with minimum $x_{\rm min}$ and tail $x^{-\alpha+1}$.
\item Correlations decay exponentially, $\left< \lambda_k(t) \lambda_k(s) \right> - \left< \lambda_k(t) \right> \left< \lambda_k(s) \right> = C e^{-|t-s|/\tau}$.
\end{enumerate}
Note that $\forall k,j$, $\lambda_k(t)$ and $\lambda_j(t)$ are independent.
Without loss of generality, we consider the case $k=1$ and generate a process in discrete time steps $\Delta t$.

The algorithm is quite simple:
\begin{enumerate}
\item Set $t=0$. Draw $\lambda_1(t)$ from the required power law.
\item Let $u \sim U(0,1)$, a uniformly distributed random variable in $(0,1)$.
\item If $u<\Delta t/\tau$, then draw $\lambda_1(t+ \Delta t)$ independently from the required power law.
Otherwise, $\lambda_1(t+ \Delta t) \leftarrow \lambda_1(t)$.
\item $t \leftarrow t+ \Delta t$. Go to 2.
\end{enumerate}
It is trivial to see that the univariate distributions are a power law. 
The (approximately) exponential correlation is due to the fact that the probability that the value of $\lambda_1(t)$ does not change after $j$ steps is $(1-\Delta t/\tau)^j$. Hence, after time $t = j \Delta t$, it is  $(1-\Delta t/\tau)^{t/\Delta t} \rightarrow_{\Delta t \to 0}  e^{-t/\tau}$.

Finally, the SIR model is simulated in steps of $\Delta t$. 
During the $i$'th step, we assume that S$\to$I transitions have an exponential distribution with constant rates $\lambda_k(i\Delta t)$.
I$\to$R transition rates equal 1.
Within each time period $\Delta t$, the SIR dynamics is modeled using the Gillespie algorithm as described in the previous section.

\section{Appendix: Self-consistent conditions for criticality}

We take advantage of a particular realization of stationary gamma processes, i.e., stochastic processes with univariate marginal gamma distribution
and an exponential auto-correlation function.
Following \cite{Wolpert2011,Wolpert2016}, if the shape parameter $\alpha$ is equal to half an integer, then one can write such a stationary gamma processes
as a stochastic differential equation.
Note that this realization is different than the AR(1) process defined in Eq. (4), i.e., it may have different higher ($\ge 3$) moments.
Specifically, for $\alpha = n/2$ and $\beta>0$, the solution of the stochastic differential equation
\begin{equation}
   dX_t = -\frac{2}{\tau} \left( X_t - \frac{n}{2\beta} \right) dt + \frac{2}{\sqrt{\beta \tau}} \sqrt{X_t} dW_t ,
\label{eq:SDE}
\end{equation}
satisfies \cite{Wolpert2011,Wolpert2016},
\begin{eqnarray}
   && X_t \sim {\rm Gamma} (n/2,\beta) \\
   && \left< X_t X_s  \right> - \left< X_t \right> \left< X_s \right> = e^{-|t-s|/\tau}
\end{eqnarray}
The Fokker-Plank equation associated with \eqref{eq:SDE} is
\begin{equation}
   \frac{\partial p(x,t)}{\partial t} = \frac{2}{\tau} \frac{\partial}{\partial x} \left[ \left( x - \frac{n}{2\beta} \right) p(x,t) \right] + \frac{2}{\beta \tau} \frac{\partial^2}{\partial x^2} \left[ x p(x,t) \right] .
\label{eq:FP0}
\end{equation}
Our goal is to model the distribution of infectivity times (here denoted $X$) in the susceptible population.
We assume that the rate in which individuals leave the state S (i.e., they become infected) is inversely proportional to the infectivity time, 
\begin{equation}
   \frac{1}{X_t} \frac{1}{ \int_0^\infty x^{-1} p(x,t) dx } .
\end{equation}
Instead of a quasi-steady state, which is difficult to define and treat analytically, we assume a steady state, which is obtained by reintroducing infected individuals back as susceptibles with infectivity times that are distributed according to the present measure, i.e. $p(x,t)$.
Note this is different than an SIS model, in which recovered individuals are also reintroduced as susceptible, but with the initial measure $p(x,0)$ (gamma here). 
Overall, the Fokker-Plank equation associated with the effective steady-state process is 
\begin{eqnarray}
  & \frac{\partial p}{\partial t} =& \frac{2}{\tau} \frac{\partial}{\partial x} \left[ \left( x - \frac{n}{2\beta} \right) p \right] + \\ 
 && \frac{2}{\beta \tau} \frac{\partial^2}{\partial x^2} \left[ x p \right] 
   - \rho \left[ \frac{1}{x} \frac{1}{ \int_0^\infty x^{-1} p(x,t) dx } -1 \right] p(x,t) ,  \nonumber
\label{eq:FP}
\end{eqnarray}
where $\rho \ge 0$ is the overall rate in which individuals are removed.
At the steady state $\partial p/\partial t=0$,
\begin{eqnarray}
  && \frac{2}{\tau} \left[ \left( x - \frac{n}{2\beta} \right) p \right]' + \frac{2}{\beta \tau} \left[ x p \right]'' \nonumber \\
 & & - \rho \left[ \frac{1}{x} \frac{1}{ \int_0^\infty x^{-1} p(x) dx } -1 \right] p(x) = 0.
\label{eq:FPsteady}
\end{eqnarray}
Taking $\rho=0$, we verify that the density of a gamma distribution with rate parameter $\beta$ and $\alpha=n/2$,
\begin{equation}
    p_0(x) = \frac{1}{\Gamma(\alpha)} \beta^{\alpha} x^{\alpha-1} e^{-\beta x} ,
\label{eq:p0}
\end{equation}
is indeed a steady state solution.
Here, $\Gamma(\cdot)$ is the gamma function.
Multiplying \eqref{eq:FPsteady} by $\tau/2$ and denoting $\epsilon=\tau \rho/2$, Eq. \eqref{eq:FPsteady} becomes,
\begin{equation}
  L_0 p(x) 
   - \epsilon \beta \left[ \frac{1}{x} \frac{1}{ \int_0^\infty x^{-1} p(x) dx } -1 \right] p(x) = 0,
\label{eq:steady}
\end{equation}
where
\begin{equation}
   L_0 f = \left[ \left( \beta x - \frac{n}{2} \right) f \right]' +  \left[ x f \right]'' ,
\end{equation}
is the forward Kolmagorov operator associated with \eqref{eq:SDE}.
Assuming $\epsilon \ll 1$, we expand $p(x)$ in an asymptotic series,
\begin{equation}
  p(x) = p_0(x) + \epsilon q(x) + O(\epsilon^2) .
\label{eq:asymptotic}
\end{equation}
Substituting into \eqref{eq:steady} and expanding in powers of $\epsilon$,
\begin{equation}
  L_0 q(x) - \beta \left[ \frac{1}{x} \frac{1}{ \int_0^\infty x^{-1} p_0 dx } -1 \right] p_0 + O(\epsilon)  = 0 ,
\label{eq:asymptotic}
\end{equation}
where we used $L_0 p_0 = 0$.
In the following, the high order $O(\epsilon)$ term is neglected.

As $\alpha=n/2$ is half an integer, we concentrate on the only relevant case with $1<\alpha<2$ and take $\alpha=3/2$.
Substituting $p_0$ into \eqref{eq:asymptotic}, we obtain a second order ordinary differential equation for the leading order perturbation term $q(x)$,
\begin{equation}
   \beta q + \left( \beta x + \frac12 \right) q' + x q'' = \frac{2}{\sqrt{\pi}} \beta^{3/2} \left( \frac{1}{\sqrt{x}} - \beta \sqrt{x} \right) e^{-\beta x} .
\label{eq:qODE}
\end{equation}
Changing variables $x=t/\beta$, we write $y(t/\beta) = q(x)$. The equation for $y$ reads,
\begin{equation}
   y + \left( t + \frac12 \right) \dot{y} + t \ddot{y} =  \frac{2}{\sqrt{\pi}}  \beta \left( \frac{1}{\sqrt{t}} - \sqrt{t} \right) e^{-t} .
\label{eq:y}
\end{equation}
This is an inhomogeneous confluent hypergeometric equation (a.k.a. Kummer's equation).
It is easily verified that $\sqrt{t} e^{-t}$ is a solution of the homogeneous equation. 
The second solution involves the imaginary error function which diverges at infinity and is therefore not normalizable.
Following the variation of parameters approach, we look for a solution in the form of 
\begin{equation}
   y(t) = \sqrt{t} e^{-t} \left[ 1 + \beta z(t) \right] .
\end{equation}
Substituting into \eqref{eq:y}, $z(t)$ satisfies
\begin{equation}
   \left( \frac{3}{2} - t \right) \dot{z} + t \ddot{z} =  \frac{2}{\sqrt{\pi}}  \left( \frac{1}{t} - 1 \right) .
\label{eq:z}
\end{equation}
Denoting $w=\dot{z}$,
\begin{equation}
   t^2 \dot{w} + t  \left( \frac{3}{2} - t \right) w =  \frac{2}{\sqrt{\pi}}  \left( 1-t \right) .
\label{eq:w}
\end{equation}
We split the right hand side into two cases
\begin{eqnarray}
   & t^2 \dot{w}_1 + \left( \frac{3}{2} - t \right) w_1 =  1 \nonumber \\
   & t^2 \dot{w}_2 + \left( \frac{3}{2} - t \right) w_2 =  t
\label{eq:w12}
\end{eqnarray}
Using Matlab's symbolic toolbox, the solutions are
\begin{eqnarray}
   && w_1(t)  = - \sqrt{\pi} t^{-3/2} e^t {\rm erfc} (\sqrt t)  \nonumber \\
   && w_2(t)  =  - \frac{1}{t} - \frac12 \sqrt{\pi} t^{-3/2} e^t {\rm erfc} (\sqrt{t}) 
\end{eqnarray}
where, erfc is the complementary error function and the constants were chosen such that $w_1$ and $w_2$ vanish in the limit $t \to \infty$.
Overall, 
\begin{eqnarray}
   w(t) =  \frac{2}{\sqrt{\pi}} \frac{1}{t} - t^{-3/2} e^t {\rm erfc} (\sqrt{t})  .
\end{eqnarray}
Integrating,
\begin{eqnarray}
   z(t) =  \frac{2}{\sqrt{\pi}} \ln t - \int_1^t  s^{-3/2} e^s {\rm erfc} (\sqrt{s}) ds + C  .
\end{eqnarray}
Substituting into $y$ and then $q$, we obtain,
\begin{eqnarray}
   &q(x) =&  \sqrt{\beta} \sqrt{x} e^{-\beta x} \left[ 1 + \beta C + \frac{2}{\sqrt{\pi}} \beta \ln (\beta x) - \right. \nonumber \\
&& \left.  \beta  \int_1^{\beta x}  s^{-3/2} e^s {\rm erfc} (\sqrt{s}) ds \right] .
\label{eq:q0}
\end{eqnarray}
The constant $C$ is determined to enforce normalization.
\begin{equation}
   1 = \int_0^\infty p(x) dx  = \int_0^\infty p_0(x) dx + \epsilon \int_0^\infty q(x) dx + O(\epsilon^2) .
\end{equation}
As $p_0$ is normalized, $\int q dx = 0$.
Denote
\begin{equation}
   h(y) = y^{1/2} e^{-y} \int_1^{y}  s^{-3/2} e^s {\rm erfc} (\sqrt{s}) ds .
\end{equation}
Expanding $h$ for small $y$, the limit $y \to 0$ exists and equals -2. In addition, the function is integrable at $\infty$.
Approximating numerically,  $\int_0^\infty h(y) dy \simeq -0.1869$.
Overall,
\begin{equation}
   C \simeq \sqrt{2\pi}  ( -0.1869 -2 + \gamma - 2 \ln 2 ) - \frac{1}{\beta},
\end{equation}
where $\gamma$ is the Euler-Mascheroni constant.
Substituting into \eqref{eq:q0}, we get that, for $x>0$,
\begin{equation}
   q(x) = p_0(x) \left[ D + \ln(\beta x) - \frac12 \sqrt{\pi}  \int_1^{\beta x}  s^{-3/2} e^s {\rm erfc} (\sqrt{s}) ds \right] 
\label{eq:q}
\end{equation}
where $D \simeq -0.2233$ and $q(0)=2 \beta$.

Here, we encounter a problem: while $q(x)$ is continuous and integrable ($\int_0^\infty q(x) dx = 0$), the correction for the average infectivity rate
$\int_0^\infty x^{-1} q(x) dx$, which also appears in \eqref{eq:steady}, diverges close to $x=0$ (because $q(0)>0$).
This seems to contradict the consistency of the asymptotic expansion in $\epsilon$, $\int x^{-1} p dt = \int x^{-1} p_0 dx + \epsilon  \int x^{-1} q dx + O(\epsilon^2)$, as the first integral is finite, while the second diverges.

In order to overcome this difficulty, we need to take into account the finite population size.
In our assumed quasi steady state, the population size is the current number of susceptible individuals, $S$.
Denote by $m$ the smallest value out of $S$ samples with density $p(x)$.
On average, we have that
${\mathbb E} [P(m)] = 1/(S+1)$, where $P(x)=\int_0^x p(t) dt$,
the cumulative distribution function of $p(x)$.
Neglecting terms of order $\epsilon$ , the leading order term is $m = \beta^{-1} S^{-2/3}$.
However, this value is random. 
Numerical tests show that, for large $S$, the standard deviation in $P(m)$ is approximately
$0.83/S$. 
Hence, the variation in $m$ cannot be neglected and needs to be also considered.

Applying a lower cut-off of all densities at $m$, we need to calculate 
\begin{equation}
   \int_m^\infty x^{-1} [p_0 (x) + \epsilon q(x) ] dx .
\end{equation}
Using known power law expansions of the error function, exponential integral ($\int t^{-1} e^{-t}$) and lower incomplete gamma function, 
\begin{eqnarray}
  && \int_m^\infty x^{-1} p_0 (x) dx = 2 \beta + O(\sqrt{m}) \\\
  && \int_m^\infty x^{-1} p_0 (x) \ln (\beta x) dx = -2 (\gamma + 2 \ln 2 ) \beta  + O(\sqrt{m}) \nonumber \\
  && \int_m^\infty x^{-1} h(\beta x) dx = - \frac{4}{\sqrt{\pi}} \left( -\gamma - \ln (\beta m) \right) + O(\sqrt{m})   \nonumber 
\end{eqnarray}
In particular, the integral over $x^{-1} q(x)$ diverges logarithmically in $S$. 
The logarithmic divergence of the last integral can also be seen from the fact that $h(x) \to -2$ as $x \to 0^+$. As a result, for $x \ll 1$, $x^{-1} h(x) \sim x^{-1}$.

Using the average value $m = \beta^{-1} S^{-2/3}$, we obtain,
\begin{equation}
  \int_m^\infty x^{-1} p(x) dx = 2 \beta  - \beta \epsilon (1.5036 + 0.5905 \ln S )  +o(1).
\label{eq:intp0}
\end{equation}
However, as explained above, this approximation is not sufficient because the
lower cutoff $m$ is used to evaluate integrals of the form $\int_m^1 x^{-1} dx$, which is not symmetric with respect to fluctuations in $m$.
In order to improve the approximation, we introduce a proportionality constant $F$
and replace \eqref{eq:intp0} with
\begin{equation}
  \int_m^\infty x^{-1} p(x) dx = 2 \beta  - \beta \epsilon (1.5036 + 0.5905 F \ln S )  +o(1).
\label{eq:intp}
\end{equation}
As notes above, $F$ describes the ratio between ${\mathbb E} [1/m]$ and  $1/{\mathbb E} [m]$,
where averages are with respect to the uniform distribution (because $h$ is approximately constant for $x \ll 1$). 
We numerically estimate the value of $F$ for the case $S=10^6$ by repeatedly sampling
$S$ samples, finding the minimum and calculating
$F={\mathbb E} [1/m]{\mathbb E} [m]$.
Because of the $1/m$ dependence, it is difficult to obtain an accurate estimate.
Using $10^5 - 10^6$ samples we find that $F$ is in the range 20-30.
Hence, below we take $F=25$.

We now need to address the consistency of the approximation for the critical quasi steady state for which it is applied.
Recall that we assume that, during the critical state, the fraction of infected individuals, $i$, is constant.
Since the recovery rate is one, the duration of the critical state is at most $1/i$, i.e., $T<1/i$.
This duration is at least $O(\ln S)$, even in exponential growth and decay regimes.
Fig. 4a shows this assumption indeed holds.
Therefore, $i<C/\ln S$.

In SIR, the fraction of infected individuals also determines the rate at which susceptible individuals vanish (i.e., become infected).
Hence, in our case, $\rho=i$.
Overall, we find that
\begin{equation}
  \epsilon  < C \frac{\tau}{2} \frac{1}{\ln S} .
\end{equation}
Hence, the expansion is consistent as long as the proportionality constant $C$ is small enough.
For example, the slope of a linear fit to Fig 4a yields $C \simeq 0.015$.

Next, we couple the steady state to the SIR dynamics. The analysis is similar to [19].
The instantaneous rate in which a susceptible individual $k$ becomes infected is $i(t)/\mu_k(t)$.
Therefore, the population-averaged rate is  $i(t) \left<1/\mu_k(t)\right>_{\rm S}$,
where $\left< \cdot \right>_{\rm S}$ denotes averaging over the susceptible individuals.
On average, the rate of change in $s(t)$ is
\begin{equation}
   \frac{ds}{dt} = -   \left<1/\mu_k(t)\right>_{\rm S} s(t) i(t).
\end{equation}
Similarly, the population-average $i$ satisfies
\begin{equation}
   \frac{di}{dt} =   \left<1/\mu_k(t)\right>_{\rm S} s(i) i(t) - i(t) = \left[ \left<1/\mu_k(t)\right>_{\rm S} s(i) - 1 \right] i(t).
\end{equation}
Thus, the reproduction rate is
\begin{equation}
   R_t = \left<1/\mu_k(t)\right>_{\rm S} s(t) = s(t) \int_0^\infty x^{-1} p(x) dx .
\label{eq:scrit}
\end{equation}

As described above, the steady state distribution of $\mu_k(t)$ has density $p(x)= p_0(x) + \epsilon q(x) + O(\epsilon^2)$ with $\epsilon=i \tau/2$.
Denoting by $i_{\rm crit}$ and $s_{\rm crit}$ the critical, steady-state values,
the reproduction rate at criticality is given by \eqref{eq:intp},
\begin{equation}
    R_t \simeq s_{\rm crit} \left[ 2 \beta  - \frac12 \beta \tau (1.5036 + 0.5905 F \ln S ) i_{\rm crit} \right],
\end{equation}
A steady state would imply the critical value $R_t=1$. 
Setting $\beta=\alpha/\left< \mu \right>=3/(2\left< \mu \right>)$ (recall that $\left< \mu \right>$ is the average infectivity time) yields,
\begin{equation}
   i_{\rm crit} = \frac{4}{\tau} \left( 1 - \frac{\left< \mu \right>}{3s_{\rm crit}} \right) \frac{1}{1.5036 + 0.5905 F \ln S}.
\label{eq:icrit}
\end{equation}
Fig.~\ref{fig5}  compares the analytic approximations to simulation results with $\alpha=3/2$, $\left< \mu \right> = 1.4$, $\tau=4$ and $N=10^6$.
We see that at the steady state (to the right of the dashed line, defined by $R_t \sim 1$),
the mean susceptibility time, the fraction of susceptible individuals and the fraction of infected individuals are well approximated by the analytical approximations \eqref{eq:intp}, \eqref{eq:scrit} and \eqref{eq:icrit}, respectively.

Next, we wish to evaluate the region in parameters space that admits the stable regime.
For simplicity, we substitute $s_{\rm crit}=1$.
As the critical $i$ is positive, we obtain a lower bound for $\left< \mu \right>$,
\begin{equation}
   \left< \mu \right><3.
\end{equation}
In addition, recall our self-consistent assumption that the perturbation in the average susceptibility rate is smaller then the population average.
This requires that $i_{\rm crit}<  2 / (\tau F \ln S)$.
Comparing with \eqref{eq:icrit}, and assuming $\ln S \gg 1$, we obtain an upper bound for $\left< \mu \right>$,
\begin{equation}
   \left< \mu \right> > 0.7.
\end{equation}
Overall, we find a consistency condition for the existence of the steady state at $\alpha=3/2$,
\begin{equation}
  0.7 < \left< \mu \right> < 3 .
\end{equation}
This result is in excellent agreement with the numerically obtained 
phase diagram Fig 2, where, with $\alpha=1.5$, critically is obtained for $0.7 < \left< \mu \right> < 2.7$.

Next, we consider the case $\alpha=2$.
Following the same protocol, we find that
\begin{eqnarray}
    && p(x) = p_0 (x) + \epsilon q(x) + O(\epsilon^2) \nonumber \\
    && p_0(x) = \beta^2 x e^{-\beta x} \nonumber \\
    && q(x) = p_0(x) (\gamma -1 + \ln \beta x)
\end{eqnarray}
Unlike the $\alpha=3/2$ case, here the average perturbed infectivity rate converges,
\begin{equation}
  \int x^{-1} [ p_0(x) + \epsilon q(x) ] dx = \beta (1-\epsilon) .
\end{equation}
As a result, assuming that $s \simeq 1$,
\begin{equation}
   R_t = \left< 1/\mu \right>_{\rm S} = \beta \left( 1 - \frac12 \tau i \right) = 1 .
\label{eq:alpha2cond}
\end{equation}
We readily see that with $\beta<1$, \eqref{eq:alpha2cond} cannot be satisfied.
With $\beta>1$, a solution may exist. However, $i=O(1)$, which implies that the duration of the steady state is independent of $N$ and therefore short.

Lastly, we take $\alpha=5/2$.
We find that
\begin{eqnarray}
    && p(x) = p_0 (x) + \epsilon q(x) + O(\epsilon^2) \nonumber \\
    && p_0(x) = \frac{4}{3\sqrt{\pi}} \beta^{5/2} x^{3/2} e^{-\beta x} \nonumber \\
    && q(x) = p_0(x) (D + \ln \beta x)
\end{eqnarray}
where $D \simeq -0.52896$.
Again, the average infectivity rate converges,
\begin{equation}
  \int x^{-1} [ p_0(x) + \epsilon q(x) ] dx = \beta (1-\epsilon) .
\end{equation}
As a result, assuming that $s \simeq 1$,
\begin{equation}
   R_t = \left< 1/\mu \right>_{\rm S} \simeq \beta \left( \frac23 - 0.7397 \epsilon \right) = 1 .
\label{eq:alpha2cond}
\end{equation}
With $\beta<3/2$, \eqref{eq:alpha2cond} cannot be satisfied.
With $\beta>3/2$, a solution may exist. However, again, $i=O(1)$, which implies that the duration of the steady state is independent of $N$ and therefore short.


\end{document}